
\hsize=35truepc\hoffset=2pc
\vsize=54truepc

\font\titlefont=cmcsc10 at 14pt
\font\namefont=cmr12
\font\placefont=cmsl12
\font\abstractfont=cmbx12
\font\secfont=cmr12

\font\twrm=cmr12         \font\twmi=cmmi12
  \font\twsy=cmsy10 at 12pt  \font\twex=cmex10 at 12pt
  \font\twit=cmti12          \font\twsl=cmsl12
  \font\twbf=cmbx12
\font\nnrm=cmr9          \font\nnmi=cmmi9
  \font\nnsy=cmsy9           \font\nnex=cmex10 at 9pt
  \font\nnit=cmti9           \font\nnsl=cmsl9
  \font\nnbf=cmbx9

\font\sxrm=cmr6          \font\sxmi=cmmi6             
  \font\sxsy=cmsy6

\def\nnpt{\def\rm{\fam0\nnrm}%
  \textfont0=\nnrm \scriptfont0=\sxrm \scriptscriptfont0=\fiverm
  \textfont1=\nnmi \scriptfont1=\sxmi \scriptscriptfont1=\fivei
  \textfont2=\nnsy \scriptfont2=\sxsy \scriptscriptfont2=\fivesy
  \textfont3=\nnex \scriptfont3=\nnex \scriptscriptfont3=\nnex
  \textfont\itfam=\nnit \def\it{\fam\itfam\nnit}%
  \textfont\slfam=\nnsl \def\sl{\fam\slfam\nnsl}%
  \textfont\bffam=\nnbf \def\bf{\fam\bffam\nnbf}%
  \normalbaselineskip=11pt
  \setbox\strutbox=\hbox{\vrule height8pt depth3pt width0pt}%
  \let\big=\nnbig \let\Big=\nnBig \let\bigg=\nnbigg \let\Bigg=\nnBigg
  \normalbaselines\rm}

\def\tnpt{\def\rm{\fam0\tenrm}%
  \textfont0=\tenrm \scriptfont0=\sevenrm \scriptscriptfont0=\fiverm
  \textfont1=\teni  \scriptfont1=\seveni  \scriptscriptfont1=\fivei
  \textfont2=\tensy \scriptfont2=\sevensy \scriptscriptfont2=\fivesy
  \textfont3=\tenex \scriptfont3=\tenex    \scriptscriptfont3=\tenex
  \textfont\itfam=\tenit \def\it{\fam\itfam\tenit}%
  \textfont\slfam=\tensl \def\sl{\fam\slfam\tensl}%
  \textfont\bffam=\tenbf \def\bf{\fam\bffam\tenbf}%
  \normalbaselineskip=12pt
  \setbox\strutbox=\hbox{\vrule height8.5pt depth3.5pt width0pt}%
  \let\big=\tenbig \let\Big=\tenBig \let\bigg=\tenbigg \let\Bigg=\tenBigg
  \normalbaselines\rm}

\def\twpt{\def\rm{\fam0\twrm}%
  \textfont0=\twrm \scriptfont0=\nnrm \scriptscriptfont0=\sevenrm
  \textfont1=\twmi \scriptfont1=\nnmi \scriptscriptfont1=\seveni
  \textfont2=\twsy \scriptfont2=\nnsy \scriptscriptfont2=\sevensy
  \textfont3=\twex \scriptfont3=\twex \scriptscriptfont3=\twex
  \textfont\itfam=\twit \def\it{\fam\itfam\twit}%
  \textfont\slfam=\twsl \def\sl{\fam\slfam\twsl}%
  \textfont\bffam=\twbf \def\bf{\fam\bffam\twbf}%
  \normalbaselineskip=14pt
  \setbox\strutbox=\hbox{\vrule height10pt depth4pt width0pt}%
  \let\big=\twbig \let\Big=\twBig \let\bigg=\twbigg \let\Bigg=\twBigg
  \normalbaselines\rm}

\catcode`\@=11
\def\nnbig#1{{\hbox{$\left#1\vbox to7.5pt{}\right.\n@space$}}}
\def\nnBig#1{{\hbox{$\left#1\vbox to10.5pt{}\right.\n@space$}}}
\def\nnbigg#1{{\hbox{$\left#1\vbox to13.5pt{}\right.\n@space$}}}
\def\nnBigg#1{{\hbox{$\left#1\vbox to16.5pt{}\right.\n@space$}}}

\def\tenbig#1{{\hbox{$\left#1\vbox to8.5pt{}\right.\n@space$}}}
\def\tenBig#1{{\hbox{$\left#1\vbox to11.5pt{}\right.\n@space$}}}
\def\tenbigg#1{{\hbox{$\left#1\vbox to14.5pt{}\right.\n@space$}}}
\def\tenBigg#1{{\hbox{$\left#1\vbox to17.5pt{}\right.\n@space$}}}

\def\twbig#1{{\hbox{$\left#1\vbox to10.5pt{}\right.\n@space$}}}
\def\twBig#1{{\hbox{$\left#1\vbox to14pt{}\right.\n@space$}}}
\def\twbigg#1{{\hbox{$\left#1\vbox to17.5pt{}\right.\n@space$}}}
\def\twBigg#1{{\hbox{$\left#1\vbox to21pt{}\right.\n@space$}}}
\catcode`\@=12

\def\center{\parindent=0pt\leftskip=1in plus 1fill\rightskip=1in plus 1fill}

\def\title#1\par{{\center\baselineskip=16pt
  \namefont{#1}\par}}

\long\def\author #1// #2// #3// #4//{{\parskip=0pt\center\namefont{#1}\medskip
  \placefont#2\par#3\par#4\par}}

\long\def\abstract#1//{%
  \centerline{\abstractfont Abstract}\medskip
  {\baselineskip=12pt\advance\leftskip by 3pc\advance\rightskip by
3pc\parindent=10pt
  \def\enspace{\kern.3em}
  \noindent #1\par}}

\def\body{\bgroup\twpt}

\def\sectitle{\center\baselineskip=14pt\lineskiplimit=1pt\secfont}

\def\beginsec#1\par{%
  \ifdim0.8\vsize<\pagetotal\ifdim\pagetotal<\pagegoal\vfil\eject\fi\else
    \removelastskip\medskip\vskip\parskip\fi
  {\sectitle #1\hfilneg\ \par}%
  \nobreak\noindent}

\def\references{\egroup\leftskip=\parindent\parindent=0pt
  \vskip0pt plus.07\vsize\penalty-250\vskip0pt plus-.07\vsize
  \removelastskip\bigskip\bigskip\vskip\parskip
  \centerline{\secfont References}\bigskip}

\newcount\refno\refno=0
\long\def\rfrnc#1//{\advance\refno by 1
  $ $\llap{\hbox to\leftskip{\the\refno.\enspace\hfil}}#1\par\medskip}
\def\PL{{\it Phys.\ Lett.}}
\def\NP{{\it Nucl.\ Phys.}}
\def\PR{{\it Phys.\ Rev.}}
\def\PRL{{\it Phys.\ Rev.\ Lett.}}

\skip\footins=0.2in
\dimen\footins=4in
\catcode`\@=11
\def\vfootnote#1{\insert\footins\bgroup\nnpt
  \interlinepenalty=\interfootnotelinepenalty
  \splittopskip=\ht\strutbox
  \splitmaxdepth=\dp\strutbox \floatingpenalty=20000
  \leftskip=0pt \rightskip=0pt \spaceskip=0pt \xspaceskip=0pt
  \setbox1=\hbox{*}\parindent=\wd1\let\enspace=\null
  \hangafter1\hangindent\parindent\textindent{#1}\footstrut
  \futurelet\next\fo@t}
\catcode`\@=12

\def\footnoterule{\kern-3pt \hrule width2truein \kern 3.6pt}

\def\norm#1{\left|#1\right|}
\pretolerance=300\tolerance=400\hyphenpenalty=100

\def\Frac#1#2{{\raise.2ex\hbox{$\scriptstyle#1$}%
  \kern-.1em\scriptstyle/
  \kern-.1em\lower.2ex\hbox{$\scriptstyle#2$}}}
\def\frac#1#2{{\raise.2ex\hbox{$\scriptscriptstyle#1$}%
  \kern-.1em\scriptscriptstyle/
  \kern-.1em\lower.2ex\hbox{$\scriptscriptstyle#2$}}}

\def\norm#1{|#1|}

\def\normm#1{\Bigl|#1\Bigr|}

\newdimen\jbarht\jbarht=.2pt
\newdimen\vgap\vgap=1pt
\newcount\shiftfactor\shiftfactor=12

\catcode`\@=11
\def\jbarout{\setbox1=\vbox{\offinterlineskip
  \dimen@=\ht0 \multiply\dimen@\shiftfactor \divide\dimen@ 100
    \hsize\wd0 \advance\hsize\dimen@
  \hbox to\hsize{\hfil
    {\multiply\dimen@-2 \advance\dimen@\wd0
    \vrule height\jbarht width\dimen@ depth0pt}%
    \hskip\dimen@}%
  \vskip\vgap\box0\par}\box1}
\def\jbar#1{\mathchoice
  {\setbox0=\hbox{$\displaystyle{#1}$}\jbarout}%
  {\setbox0=\hbox{$\textstyle{#1}$}\jbarout}%
  {\setbox0=\hbox{$\scriptstyle{#1}$}\jbarout}%
  {\setbox0=\hbox{$\scriptscriptstyle{#1}$}\jbarout}}
\catcode`\@=12


\def\cut#1{\setbox0=\hbox{$#1$}\setbox1=\hbox to \wd0{\hss{\it/\/}\hss}
  \box1\hskip-\wd0\box0}                   


\def\simlt{\mathop{\lower.4ex\hbox{$\buildrel<\over\sim$}}}
\def\simgt{\mathop{\lower.4ex\hbox{$\buildrel>\over\sim$}}}

\font\secfont=cmbx12
\font\titlefont=cmbx12 at 14pt
\font\submitfont=cmti12
\font\MIUfont=cmr12
\vsize=52truepc\voffset=1pc
\hsize=37truepc\hoffset=1pc
\pageno=1
\footline{\ifnum\pageno>1\hss\lower.25in\hbox{\tenrm\folio}\hss\fi}

\widowpenalty=1000
\advance\baselineskip by 2pt

\advance\baselineskip by 2pt
\null\vskip-2\baselineskip
\rightline{\submitfont Submitted to Physical Review Letters}
\vskip\baselineskip
\rightline{\MIUfont MIU-THP-92/62}
\rightline{\MIUfont September, 1992}
\vskip.5in
\title{\titlefont COBE and SUSY}
\vskip.2in
\author
Lawrence Connors, Ashley J. Deans and John S. Hagelin//
Department of Physics//
Maharishi International University//
Fairfield, Iowa 52557//

\vskip.4in

\abstract\advance\leftskip by 20pt\advance\rightskip by 20pt
We show that supersymmetry automatically leads to density fluctuations $\Delta
T/T\sim6\times10^{-6}$ in agreement with the recent COBE measurement.
\par//
\vskip.4in

\body
\parindent=25pt
\parskip\medskipamount

\beginsec 1. Introduction

The cosmic microwave background radiation (CMBR) is of great interest to
cosmologists because it is thought to faithfully reflect conditions in the
early universe when electrons recombined with nuclei to form neutral atoms
[1].  After this event, which happened when the universe was about 100,000
years old, photons stopped scattering off matter and simply redshifted to the
2.735K blackbody spectrum observed today.  Perhaps the most interesting feature
of the CMBR is its very small degree of temperature fluctuations $\Delta
T/T\sim6\times10^{-6}$, which has recently been measured for the first time by
the COBE satellite [2].  This measurement has given strong support to the idea
that the CMBR fluctuations $\Delta T/T$ derive from the same source as the
matter fluctuations $\delta \rho/\rho\sim10^{-4}$ which were necessary to seed
galaxy formation.  In the most popular theories that incorporate this concept
[3], both $\Delta T/T$ and $\delta\rho/\rho$ are produced by perturbations of a
scalar quantum field $\phi$ during a primordial inflationary period.
Typical scenarios using cold dark matter predict that
$\delta\rho/\rho\sim(10-100)\Delta T/T$ [1], which is consistent with COBE's
measurement.

If the Lagrangian responsible for inflation contains the scalar field $\phi$ in
the generic form $\lambda\phi^4$, then a given value of $\delta\rho/\rho$
requires a value of
$\lambda={\frac32}(\frac{3\pi}8)^2N^{-3}(\delta\rho/\rho)^2$, where $N$ is the
number of e-foldings before the end of inflation that the relevant
perturbations ``leave the horizon" of the inflationary universe.  For
$\delta\rho/\rho=10^{-4}$ and $N=60$ (the number required to insure the
flatness and homogeneity of the universe today) this demands a value of
$\lambda\sim10^{-13}$; in realistic models [4,1] a lower $N$ is usually
relevant, but still a $\lambda\sim10^{-11}-10^{-12}$ is required.  The main
difficulty in implementing inflationary models has always been the need to
motivate this extremely small dimensionless parameter
$\lambda\sim10^{-11}-10^{-12}$.  In non-supersymmetric theories, this has
been achieved by unnatural fine-tuning of parameters and/or the introduction
of {\sl ad hoc}, non-renormalizable potentials for observable or
hidden-sector scalar fields [5].

We show in this paper that all realistic supersymmetric theories (i.e., those
containing an electron) naturally predict this magnitude of density
fluctuations.  This is because the supersymmetric scalar potential
automatically includes a ``quasi-flat" $\lambda\phi^4$ direction with
$$\lambda=(\lambda^e)^2=\frac12g^2_2(m_e/m_W)^2(1+\langle h^o\rangle^2/\langle
\jbar h^o\rangle^2)\sim10^{-11},\eqno{(1)}$$
where $\lambda^e$ is the electron Yukawa coupling.  It is worth emphasizing
that this parameter naturally arises from the supersymmetric relationship
$\lambda=(\lambda^e)^2$ between the quartic scalar potential and the
superpotential, which contains the somewhat small electron Yukawa coupling
$\lambda^e\sim10^{-5}$.  Moreover, this small quartic scalar coupling is
multiplicatively renormalized, and thus remains small to all orders in
perturbation theory.

[Another related useful feature of supersymmetry is the likely presence [6] of
stable, weakly-interacting massive particles (WIMPs), which provide highly
plausible cold dark matter candidates [7].  Such dark matter is essential for
the
initial density fluctuations described above to begin to seed galaxy formation
prior to recombination, thereby giving these fluctuations sufficient time to
grow [1].]

In this paper we argue that the primordial density fluctuations arose
during a period of chaotic inflation [8], which seems virtually inescapable in
supersymmetric theories due to the large number of scalar fields (squarks,
sleptons and higgses) present in such models.  At timescales $\sim M^{-1}_P$
characteristic of the early universe, the uncertainty principle makes it highly
unlikely that all of these fields $\phi_i$ would be localized initially within
a
distance $\langle\phi_i\rangle<M_P$ of the origin.  This makes it almost
inevitable that one would obtain through chaotic inflation the $>$ 60
e-foldings of expansion needed for a viable inflationary scenario.

We show explicitly how this chaotic inflation leads to density fluctuations
$\delta\rho/\rho\sim 10^{-4}$ $(\Delta T/T\sim 10^{-5}-10^{-6})$ consistent
with
COBE, and under plausible assumptions to a baryon asymmetry $n_B/s\sim
10^{-10}$.

\beginsec 2.  The SUSY Scalar Potential

In order to identify the $\lambda\phi^4$ direction whose $\lambda$ is described
by Eq.~(1), we recall that the scalar potential V of a supersymmetric
Lagrangian
is composed of both D-terms and F-terms, with
$$D=\sum_a\normm{{g_a\over\sqrt 2}\sum_{kl}\phi^*_k
    T^a_{kl}\phi_l}^2 \hskip.1in{\rm and }\hskip.1in
  F=\sum_k \normm{{\partial W\over
   \partial\phi_k}}^2,\eqno{(2)}$$
where $\phi_k$ ranges over all scalar fields in the theory, the $T^a$ are the
symmetry generators, and W is the superpotential.  For definiteness, we will
use
the superpotential of the minimal supersymmetric Standard Model (MSSM) [9]:
$$W=\lambda{^i_1}\epsilon_{rs}Q_{i\alpha r}d^c_{i\alpha}h_s
   +\lambda{^{ij}_2}\epsilon_{rs}Q_{i\alpha r}u^c_{j\alpha}\jbar h_s
   +\lambda^i_3\epsilon_{rs}L_{ir}l^c_ih_s \eqno{(3)}$$
[$i,j=$ family indices; $r,s=SU(2)$; $\alpha=SU(3)$], since at least these
terms
will be present in the superpotential of any other realistic supersymmetric
theory (e.g., SUSY GUTS, etc.).  For convenience, we
have chosen a basis in which both $\lambda_1$ and $\lambda_3$ are diagonal; in
particular we have $\lambda^1_3=\lambda^e$.

The coefficients of the D-terms (i.e., the $g_a$) are all presumably {\it
O}(1).  It
is the F-terms which convert the products of possibly-small Yukawa couplings
$\lambda^{ij}_k$ in W into coefficients of quartic terms.  A simple
one-parameter example of a quasi-flat direction is
$$\tilde e=\sqrt2a;\hskip.2in\tilde e^c=\tilde c_{\alpha}=\tilde
d^c_{\jbar\alpha}
  =\tilde\nu_\mu=a, \eqno{(4)}$$
where $a$ is any complex number and all other fields are zero.  (Tildes denote
scalar partners of Standard Model quarks and leptons.)  This direction is
D-flat, and it is also F-flat except for the extremely small term
$\norm{\partial W/\partial h_1}^2=\norm{\lambda^e\tilde e\tilde
e^c}^2=\norm{\lambda^e}^2 \norm{a}^4$.  A two-parameter generalization of
this direction is
$$\tilde e=\sqrt{{\norm a}^2+{\norm b}^2}e^{i\phi};
\hskip.2in\tilde e^c=\tilde\nu_\mu=a;
\hskip.2in\tilde c_\alpha=\tilde d^c_{\jbar\alpha}=b.
\eqno {(5)}$$
There are many other such quasi-flat directions, making it all the more likely
that at least one such direction will have initial conditions
$\norm\phi\simgt$ few $\times M_P$, generating the required $>60$
e-foldings of quasi-flat chaotic inflation.

\beginsec 3.  Generation of Density Perturbations

Following the principles of chaotic inflation, we expect that
all scalar fields at the Planck time $t=M_P^{-1}$ will have values {\it
O}$(M_P)$.
For a random set of initial conditions $\phi^I_i$, those directions in field
space corresponding to F and D non-flat directions evolve quickly in a
$\lambda\phi^4$-type potential:
$$\ddot\phi_i+3H\dot\phi_i=-{\partial
V\over\partial\phi_i^*},\hbox{\hskip.25in}
  H^2={8\pi\rho\over3M_P^2}.\eqno{(6)}$$

The directions in field space with the steepest slopes evolve first, moving in
toward the origin or into nearby valleys if the potential $V(\phi)$ contains
flat directions. This initial phase of evolution in a generic $\lambda\phi^4$
potential contributes $N\sim\pi({\phi^I/M_P})^2$ e-foldings of
inflation [8], where $\phi^I$ characterizes the magnitude of the largest
initial field v.e.v.'s.  [For this discussion, we assume that the scalar fields
$\phi_i$ evolve according to their classical equations of motion [Eq.~(6)].
Although we expect quantum fluctuations $\langle\phi^2_i\rangle$ to play an
important role during the initial phase of chaotic inflation, our
semi-classical approximation (6) becomes a good approximation by the onset of
quasi-flat inflation.]

The last of the F and D non-flat directions to evolve is the quasi-flat
direction
$\Phi_{QF}$ with its quartic self-coupling
$\lambda=\norm{\lambda_3^e}^2\sim10^{-11}$ [10].  With an initial
characteristic size $\Phi^I_{QF}\simgt4M_P$ for any one of the many quasi-flat
directions, this direction would produce
$N_{QF}\sim\pi(\Phi^I_{QF}/M_P)^2\simgt50$ e-foldings of inflation during the
final quasi-flat period.  Standard theories [1] predict that
the density perturbations which influenced galactic structure formation on
distance scales between 1--1000 Mpc had their origin approximately 45-50
e-foldings before the end of inflation. This falls within the above quasi-flat
inflationary period and therefore guarantees a value of
$\delta\rho/\rho\sim10^{-4}$ in accord with observation.

\beginsec 4. Automatic Baryogenesis

Most quasi-flat directions in the scalar potential belong to families
characterized by several parameters, where the potential V may be independent
of
one or more of these parameters. We see this in the following nine-parameter
quasi-flat direction (where radicals are understood to be multiplied by
arbitrary
phases):
\bgroup\tnpt
$$\def\na{\hbox{---}}
  \def\struta{\mathord{\vrule height0pt depth7pt width0pt}}
  \def\strutb{\mathord{\vrule height0pt depth13pt width0pt}}
  \def\strutc{\mathord{\vrule height0pt depth16pt width0pt}}
  \def\gap#1 {&height#1pt\cr}
  \vcenter{\offinterlineskip\tabskip=0pt\halign{\tabskip=2pt
  \hfil#\hfil&\vrule#\tabskip=10pt&&\hfil$\displaystyle{#}$\hfil
    \tabskip=0pt&#\tabskip10pt\cr
  Gen&&\tilde Q_1&&\tilde Q_2&&\tilde{\jbar U}&&\tilde{\jbar D}
     &&\tilde L_1&&\tilde L_2&&\tilde L^c\cr\gap2
  \noalign{\hrule}\gap15
  1&&a_\alpha&&\na&&b_{\jbar \gamma}&&c_{\jbar \beta}
     &&\na&&d&&e\cr\gap15
  2&&\sqrt{{\norm d}^2+{\norm h}^2\atop-{\norm e}^2-{\norm a}^2}\strutb_\alpha
     &&\na&&f_{\jbar \gamma}
     &&g_{\jbar \beta}&&\na&&h&&\na\cr\gap12
  3&&\na&&\sqrt{{\norm m}^2-{\norm e}^2}\,\struta_\beta
     &&\sqrt{\matrix{{\norm e}^2+{\norm c}^2\cr +{\norm g}^2-{\norm m}^2\cr
       -{\norm b}^2-{\norm f}^2\cr}}
       \strutc_{\jbar\gamma}
     &&\sqrt{\matrix{{\norm d}^2+{\norm h}^2\cr +{\norm c}^2+{\norm g}^2\cr
       -{\norm m}^2}}
       \strutc_{\jbar \alpha}&&m&&\na&&\na\cr
}}\eqno{(7)}$$
\egroup
In this example the potential V is again given simply by
$\norm{\lambda^ede}^2$. As $e$ and/or $d$ fall toward zero, the other
parameters
generally remain at high values [11]. In other words the quasi-flat direction
rolls down
into an eight-parameter valley at a large distance from the origin.

A useful feature of this eight-parameter valley is that it gives non-zero
expectation value to the following baryon number-violating operator:
$$O_{BX}=\biggl(\sum_{gen}\tilde Q\tilde{\jbar U}^*\biggr)
         \biggl(\sum_{gen}\tilde L\tilde{\jbar D}^*\biggr)\eqno{(8)}$$
(the antisymmetric  $SU(2)$ and $SU(3)$ indices have been suppressed).
Such flat directions have been analyzed in detail by Affleck and Dine [12], who
showed that they give rise to a baryon asymmetry $n_B/s\sim1$ under the
low-temperature conditions likely to prevail after inflation [13]. (A
requirement of their analysis was the lifting of the degeneracy of the flat
direction through an $m^2\phi^2$ potential induced by soft-supersymmetry
breaking.) Such baryogenesis is particularly efficient when the fields
$\Phi_{AD}$ in $O_{BX}$ have values in the region $M_{GUT}<\Phi_{AD}<M_P$.
Affleck and Dine argued that such v.e.v.'s could be generated along a flat
direction by starting at the origin and then being driven up by quantum
fluctuations in the early universe.  In our model, these high v.e.v.'s are the
natural endpoint of evolution from the quasi-flat direction (7), providing (we
feel) a much better motivated lead-in to the elegant Affleck-Dine baryogenesis
scenario.

In the most successful and plausible supersymmetric GUT, Flipped $SU(5)$ [14],
the
GUT symmetry is broken by the v.e.v. of a weakly-coupled field $\Phi$, which
corresponds to a flat direction in the scalar potential. We have shown
elsewhere [4] that the slow decay of the oscillations of this weakly-coupled
field about the GUT minimum in a soft supersymmetry-breaking $m^2\Phi^2$
potential generates an entropy density $s$ that is sufficient to dilute Affleck
and Dine's problematic $n_B/s\sim1$ down to today's observed
$n_B/s\simeq10^{-10}$. Thus automatic baryogenesis can be a natural side
benefit of
the above mechanism for generating density perturbations.

\beginsec 5. Conclusion

We have demonstrated that realistic supersymmetric theories contain many
quasi-flat $\lambda\phi^4$ directions with $\lambda=(\lambda^e)^2\sim10^{-11}$
which naturally lead to density perturbations $\delta\rho/\rho\sim10^{-4}$
consistent with the recent COBE measurement $\Delta T/T\sim6\times10^{-6}$. In
addition, we have shown that many of these quasi-flat directions flow directly
into Affleck-Dine-type flat valleys at a distance $\simgt M_P$ from the
origin. In a supersymmetric GUT this will lead to Affleck-Dine baryosynthesis
resulting in $n_B/s\sim1$. If this GUT is broken by a v.e.v. along a flat
direction [as in Flipped $SU(5)$], this large baryon asymmetry is naturally
diluted by the decaying oscillations of this weakly-coupled flat direction.

We conclude that supersymmetric theories automatically explain a number of
cosmological features, including $\delta\rho/\rho\sim10^{-4}$, that they have
not yet been credited for.

This research was supported in part by the National Science
Foundation, grant no. PHY-9118320.

\references

\rfrnc E.W. Kolb and M.S. Turner, {\it The Early Universe}, Addison-Wesley,
Redwood
City, CA, 1990.//

\rfrnc G. Smoot et al., {\it Ap. J. Lett.}\/ {\bf 396} (1992) L1; E.L. Wright
et al., {\it ibid} L13.//

\rfrnc A. Guth and S. Y. Pi, \PRL\/ {\bf 49} (1982) 1110; S. Hawking, \PL\/
{\bf 115B}
(1982) 295; A.A. Starobinskii, \PL\/ {\bf 117B} (1982) 175; J.M. Bardeen, P.J.
Steinhardt
and M.S. Turner, \PR\/ {\bf D28} (1983) 679.//

\rfrnc L. Connors, A.J. Deans and J.S. Hagelin, \PL\/ {\bf 220B} (1989) 368; L.
Connors, Ph.D. thesis, Maharishi International University.//

\rfrnc For a review, see K.A. Olive, {\it Phys. Rep.} {\bf 190} (1990) 181,
especially Sec. 5.//

\rfrnc J. Ellis, J.S. Hagelin, D.V. Nanopoulos, K.A. Olive and M. Srednicki,
\NP\/ {\bf B238} (1984) 453.//

\rfrnc M.J. West, J.V. Villumsen and A. Dekel, {\it Ap. J.}\/ {\bf 369} (1991)
287.//

\rfrnc A.D. Linde, {\it Particle Physics and Inflationary Cosmology}, Harwood
Academic Publishers, NY, 1990; A.D. Linde, {\it Rep. Prog. Phys.} {\bf 47}
(1984)
925.//

\rfrnc See for example, J. Rosiek, \PR\/ {\bf D41} (1990) 3464.//

\rfrnc The duration of $\lambda\phi^4$ chaotic inflation is proportional to
$\lambda^{-1/2}$; e.g., the ``electon slope inflation" should last longer than
the
``muon slope inflation" by a factor $\lambda^\mu/\lambda^e=m_\mu/m_e\sim200$.
Calculations show that $(\lambda^e)^2\Phi^4_{QF}$ inflation lasts
$\sim10^6M^{-1}_P$ [4].//

\rfrnc If only one of the parameters $d$ and $e$ falls to zero (for example,
$e$), then
we effectively have a case of $V={1\over2} m^2\phi^2$ chaotic inflation, with
$m^2=2\lambda d^2\sim10^{11}M^2_P$. The density perturbations will then be [8]
$\delta\rho/\rho=8/(3\sqrt{3\pi})\hskip.05in(m/M_P)N\sim50\times10^{-5.5}\sim10^{-4}$,
similar to above. (Here $N\sim50$ is the number of e-foldings before the end of
inflation when the perturbations leave the horizon.)//

\rfrnc I. Affleck and M. Dine, \NP\/ {\bf B249} (1985) 361.//

\rfrnc The sample flat direction given in their paper was equivalent to our
quasi-flat direction (7) with all parameters equal to zero except $a=h$ and
$b=g$.//

\rfrnc I. Antoniadis, J. Ellis, J.S. Hagelin and D.V. Nanopoulos, \PL\/ {\bf
B194} (1987) 231.//

\end